# Towards an interoperable information infrastructure providing decision support for genomic medicine

Matthias SAMWALD[a,b,1], Holger STENZHORN[c], Michel DUMONTIER[d], M. Scott MARSHALL[e,f], Joanne LUCIANO[g,h], and Klaus-Peter ADLASSNIG[i]

[a] *Section for Medical Expert and Knowledge-Based Systems, Center for Medical Statistics, Informatics, and Intelligent Systems, Medical University of Vienna, Austria*
[b] *Institute of Software Technology and Interactive Systems, Technical University of Vienna, Austria*
[c] *Department of Pediatric Oncology and Hematology, Saarland University Hospital, Germany*
[d] *Department of Biology, Institute of Biochemistry, School of Computer Science, Carleton University, Canada*
[e] *Informatics Institute, University of Amsterdam, The Netherlands*
[f] *Department of Medical Statistics and Bioinformatics, Leiden University Medical Center, The Netherlands*
[g] *Rensselaer Polytechnic Institute, USA*
[h] *Predictive Medicine, Inc., USA*
[i] *Medexter Healthcare Gmbh, Austria*

**Abstract.** Genetic dispositions play a major role in individual disease risk and treatment response. Genomic medicine, in which medical decisions are refined by genetic information of particular patients, is becoming increasingly important. Here we describe our work and future visions around the creation of a distributed infrastructure for pharmacogenetic data and medical decision support, based on industry standards such as the Web Ontology Language (OWL) and the Arden Syntax.

## Introduction

There is growing consensus in the medical and pharmaceutical community that further progress in the development of new therapies will necessitate a fundamental change in medical practice: away from broadly defined disease concepts and therapeutic regimes, and towards a fine-tuned evidence-based, personalized medicine. Genomic medicine is an important component of personalized medicine, and refers to a system in which medical decisions are refined by combining medical history with current physiological indicators against a genetic background for a particular patient [1]. Since genetics plays a major role in determining the response to a broad range of therapeutic treatments, the appropriate use of this pharmacogenetic information for guiding treatment decisions has the potential to improve the efficacy of treatments and reduce the incidence of adverse drug events.

---
[1] Corresponding Author.

While nearly one fourth of all outpatients in the US received one or more drugs for which pharmacogenetic knowledge is available [2], it is still not common that pharmacogenetic findings are used in medical practice. Doctors are usually not specifically trained in genomic medicine, the cost-benefit trade-off of genetic testing is often unclear, and there is not enough time to incorporate potentially complex pharmacogenetic reasoning in routine medical decision making.

Therefore, the development of decision support systems capable of handling pharmacogenetic data is clearly essential to the realization of personalized medicine. These systems need to provide accurate and timely reminders and decision support tailored to each individual patient, drug and therapeutic regime. However, creators of decision support systems for genomic medicine face the challenge of working with highly heterogeneous information concerning the relationship between genetics and drug responses based on limited trials. They need to deal with distributed, incomplete and possibly contradictory information.

Here, we describe our ongoing work and future visions of employing information technologies to address this problem and towards 1) seamless integration of relevant pharmacogenetic data in a distributed setting, 2) the exploitation of clinically relevant pharmacogenetic knowledge in clinical decision support and 3) the design and dissemination of clinical decision support systems that improve the quality of health care delivery.

## 1. Methods

### 1.1. Data sources

Several relevant data sources have already become available in an open, interlinked format, or will be made available soon. We, together with other participants of the *Health Care and Life Science Interest Group* [3] of the World Wide Web Consortium (W3C, [4]), worked on making several relevant datasets accessible in RDF/OWL [5] format. The extraction and conversion of additional relevant datasets such as the Pharmacogenomics Knowledge Base (PharmGKB [6]), Drugbank [7], Online Mendelian Inheritance in Man (OMIM [8]), dbSNP [9] or SNPedia [10] is currently ongoing.

In addition to manually curated data, natural language processing has been successfully used to identify pharmacogenomic information, such as gene-drug-disease relationships [11] or descriptions of new molecular diagnostics [12].

Organisations dedicated to reviewing current evidence and publishing recommendations about pharmacogenetics have emerged. For example, the Clinical Pharmacogenetics Implementation Consortium (CPIC) was recently initiated in the context of the PharmGKB. The CPIC members create, curate, review, and update written summaries and recommendations for implementing specific pharmacogenetic practices. Levels of evidence and strength of recommendations are documented. Another example of such an organisation is the Evaluation of Genomic Applications in Practice and Prevention initiative (EGAPP [13]). The text-based recommendations provided by such initiatives can be easily formalized as rules for clinical decision support.

*1.2. Enabling data integration and semantic interoperability*

Ontologies help improve interoperability and data consistency. Several ontologies relevant to pharmacogenetics have become available in recent years. The *Translational Medicine Ontology* (TMO, [14]) provides a foundation upon which chemical, genomic and proteomic data can be harmonized and linked to disease, treatments and electronic health records. The *Suggested Ontology for Pharmacogenomics* (SO-PHARM, [15]) was the first to demonstrate how pharmacogenomic knowledge can be captured based on the Open Biomedical Ontologies (OBO) resources. The *Sequence Ontology* aims to describe the features and attributes of biological sequences [16]. It holds terms and relations of value for describing genetic variation including single nucleotide polymorphisms (SNPs) at the sequence level.

Our work is guided by international standardisation efforts, and we also participate in standardisation activities. The most important standardisation organizations in this context are Health Level 7 (HL7 [17]); and the World Wide Web Consortium (W3C), which develops standards for large-scale, distributed data integration and access.

A number of developments in the pharmaceutical domain should help to drive the practice of applying standards for interoperable information systems. The European FP7 Innovative Medicines Initiative (IMI) grants, with matching sponsorship from pharmaceutical companies, have created several projects which need interoperable information systems in order to share results and information across several IMI projects that cover domains including drug discovery, electronic patient records, clinical trials, quantitative modeling, and tissue banking. Participants of the IMI projects include many academic and pharmaceutical partners, as well as participants in the EU Biobanking and Biomolecular Resources Research Infrastructure (BBMRI), which aims to improve access to biological resources required for health-related research and development.

*1.3. Creating decision support systems*

Rule-based systems are useful for creating pharmacogenetic decision support systems [18]. We are exploring the use of standards-based rule frameworks such as the Arden Syntax [19] for this task. Arden Syntax is an HL7 standard that specifies various aspects of medical logic representation, including mechanisms for triggering rules based on certain conditions, retrieval of data from medical information systems and generating conclusions from input data.

Since current findings about the relationships between genetic variability, diseases and treatment responses are often vague and contradictory, the use of classical rule engines can be augmented by fuzzy and probabilistic reasoning and consistency checking. This is being addressed by recently created systems such as Fuzzy Arden Syntax [20] or the probabilistic OWL reasoner Pronto [21].

To ensure that these developments have a real impact on clinical practice, they will be complemented by extensive collaboration with clinical practitioners and international stakeholders. Key factors for successful deployment of decision support systems have been described in the literature [22]. Based on these findings, the systems we envision need to be directly connected to hospital information systems, seamlessly integrated into existing workflows and able to handle information from electronic patient records and clinical laboratories.

## 2. Preliminary results and discussion

The Medical University of Vienna together with the Vienna General Hospital are currently finalizing the establishment of an informatics platform for integrating clinical data with genomic data, as well as providing clinical decision support based on the Arden Syntax (Fig. 1).

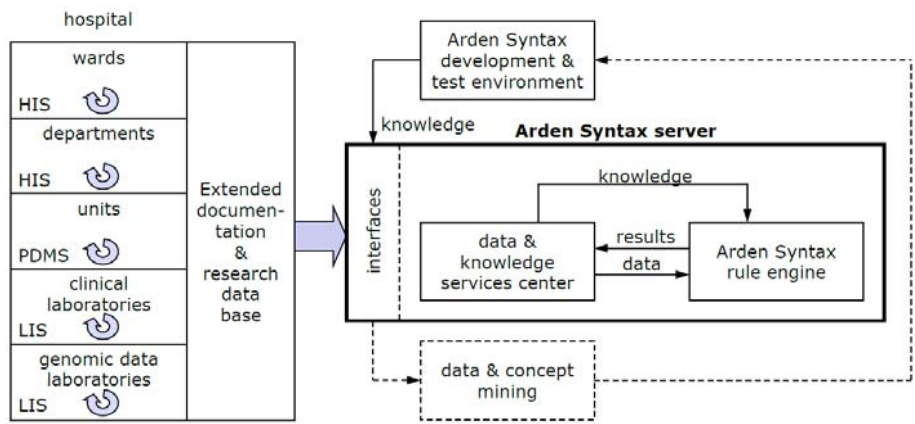

**Figure 1.** The Arden Syntax-based decision support infrastructure at the Vienna General Hospital. The Medical University of Vienna is working in close collaboration with the Vienna General Hospital to implement a new hospital information system that is used for patient care, clinical research and documentation. The clinical decision support system is part of the new hospital information system. Clinical and genomic data of patients can be transferred to the service-enabled clinical decision support server.

Another relevant development is the European integrated project *p-medicine*, which started recently. It focuses on the transformation from reactive to preventive medicine and a novel systems approach on integrated diagnosis, treatment and prevention in individuals. Within the project, an open, standards-compliant and modular framework of tools and services is being developed to enable efficient, secure sharing and handling of personalized data and in-silico models. Some important aspects are privacy, non-discrimination, and access policies to maximize patient protection and benefit. The tools are being validated within concrete, advanced clinical research settings: Pilot cancer trials have been selected on clear research objectives to emphasize the need for multilevel data integration. One specific task in p-medicine to provide capabilities to communicate directly with existing clinical trial and hospital information systems via push and synchronization services. These services are being implemented based on existing standards, such as HL7, SNOMED CT, *International Classification of Diseases and Related Health Problems, Tenth Revision* (ICD-10), specifications of the *Clinical Data Interchange Standards Consortium* (CDISC) and *Logical Observation Identifiers Names and Codes* (LOINC) to overcome the inherent heterogeneity of those systems.

We expect the research programme outlined in this paper to have several implications for clinical practice, such as improving the translation of basic pharmacogenomic findings into clinical practice, increasing the deployment of automated clinical reminders based on patient characteristics and, ultimately, improving the quality of treatments.